\documentclass[journal,twoside,web]{ieeecolor}
\usepackage{tmi}
\usepackage{cite}
\usepackage{amsmath,amssymb,amsfonts}
\usepackage{algorithmic}
\usepackage{graphicx}
\usepackage{textcomp}

\usepackage{url}
\usepackage{color}
\usepackage{subfigure}
\usepackage{graphicx}
\usepackage{multirow}
\usepackage{booktabs} 
\usepackage[marginal]{footmisc} 
\usepackage{tipa} 
\usepackage{float} 
\usepackage{placeins}
\usepackage{bm} 
\usepackage{array} 
\usepackage{enumerate}
\usepackage{hyperref}
\usepackage{pifont} 
\usepackage{subfloat}
\usepackage{subfigure}
\usepackage{amsmath,bm}

\usepackage{diagbox}
\usepackage{booktabs}
\usepackage{amssymb}
\usepackage{balance} 
\usepackage{bbm} 
\usepackage[dvipsnames]{xcolor} 
\definecolor{instance}{RGB}{252, 176, 8}
\definecolor{semantic}{RGB}{231, 71, 105}

\usepackage{amsmath}

\def\BibTeX{{\rm B\kern-.05em{\sc i\kern-.025em b}\kern-.08em
    T\kern-.1667em\lower.7ex\hbox{E}\kern-.125emX}}
\begin{document}

\title{HER-Seg: Holistically Efficient Segmentation for High-Resolution Medical Images}
\author{Qing Xu, Zhenye Lou, Chenxin Li, Yue Li, Xiangjian He, \IEEEmembership{Senior Member, IEEE}, Fiseha Berhanu Tesema, Rong Qu, \IEEEmembership{Senior Member, IEEE}, Wenting Duan, Zhen Chen, \IEEEmembership{Member, IEEE}
\thanks{\quad This work is partially supported by the Yongjiang Technology Innovation Project (2022A-097-G), and National Natural Science Foundation of China Grant (UNNC: B0166). (Corresponding author: Xiangjian He) \\ Q. Xu, Y. Li, X. He, F. Tesem, are with the School of Computer Science, University of Nottingham Ningbo China, China, and with University of Nottingham, UK (e-mail: sean.he@nottingham.edu.cn). \\ Z. Lou is with Sichuan University Pittsburgh Institute, Sichuan University, China (e-mail: leonlou0921@gmail.com). \\ C. Li is with The Chinese University of Hong Kong, Hong Kong SAR (e-mail: chenxinli@link.cuhk.edu.hk). \\ R. Qu is with the School of Computer Science, University of Nottingham, UK. (email: rong.qu@nottingham.ac.uk). \\ W. Duan is with the School of Computer Science, University of Lincoln, UK. (email: wduan@lincoln.ac.uk). \\ Z. Chen is with HKISI, Chinese Academy of Sciences, Hong Kong SAR (e-mail: zchen.francis@gmail.com).
}}

\maketitle
\begin{abstract}
High-resolution segmentation is critical for precise disease diagnosis by extracting fine-grained morphological details. Existing hierarchical encoder-decoder frameworks have demonstrated remarkable adaptability across diverse medical segmentation tasks. While beneficial, they usually require the huge computation and memory cost when handling large-size segmentation, which limits their applications in foundation model building and real-world clinical scenarios. To address this limitation, we propose a holistically efficient framework for high-resolution medical image segmentation, called HER-Seg. Specifically, we first devise a computation-efficient image encoder (CE-Encoder) to model long-range dependencies with linear complexity while maintaining sufficient representations. In particular, we introduce the dual-gated linear attention (DLA) mechanism to perform cascaded token filtering, selectively retaining important tokens while ignoring irrelevant ones to enhance attention computation efficiency. Then, we introduce a memory-efficient mask decoder (ME-Decoder) to eliminate the demand for the hierarchical structure by leveraging cross-scale segmentation decoding. Extensive experiments reveal that HER-Seg outperforms state-of-the-arts in high-resolution medical 2D, 3D and video segmentation tasks. In particular, our HER-Seg requires only 0.59GB training GPU memory and 9.39G inference FLOPs per 1024$\times$1024 image, demonstrating superior memory and computation efficiency. The code is available at \url{https://github.com/xq141839/HER-Seg}.
\end{abstract}


\section{Introduction}
\label{sec:introduction}

\IEEEPARstart{H}{igh-resolution} medical images play a pivotal role in identifying microstructure information of tissue and organs as well as tiny lesion change from diverse modalities, \textit{e.g.,} dermoscopy, fundus, and microscopy, advancing clinical applications in terms of precise disease diagnosis \cite{wang2025serp}. Moreover, mobile devices have demonstrated exceptional potential in enabling fast and more accessible medical image segmentation and disease diagnosis \cite{wang2024repvit}. Such computational resource limitations pose a significant challenge to developing high-resolution medical image segmentation frameworks with the necessary efficiency. The classical U-shape encoder-decoder framework \cite{ronneberger2015u} with convolutional neural network (CNN) demonstrates superior adaptability for diverse medical segmentation tasks due to its outstanding inductive bias capabilities. Despite the advantage, existing CNN-based UNet variants \cite{isensee2021nnu, ibtehaz2023acc, zhu2024selfreg, xu2024lb} are difficult to capture global contexts due to the constraints of local receptive fields, affecting
their segmentation accuracy when handling large-size medical images.

\begin{figure}[!t]
  \centering
  \includegraphics[width=0.95\linewidth]{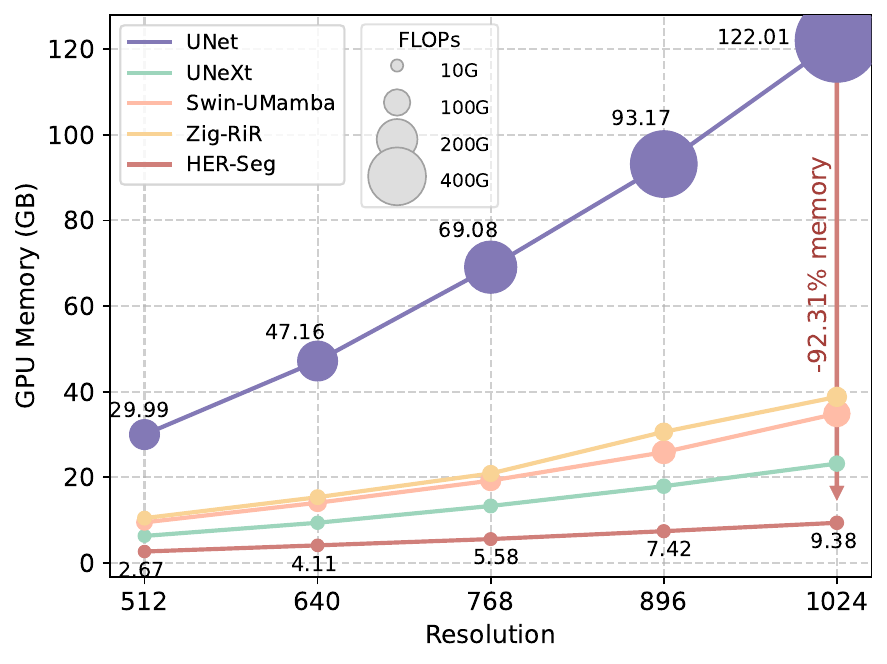}
  \caption{Comparison on training memory and FLOPs cost. Our HER-Seg demonstrates the lowest FLOPs and respectively reduces GPU memory cost by 92.31\% and 59.59\% compared to standard UNet \cite{ronneberger2015u} and efficient UNeXt \cite{valanarasu2022unext} when performing a standard training setting (e.g., 16 batches) with high-resolution medical images of 1024$\times$1024.}
  \label{fig:intro_memory}
\end{figure}

\begin{figure*}[!t]
  \centering
  \includegraphics[width=0.95\linewidth]{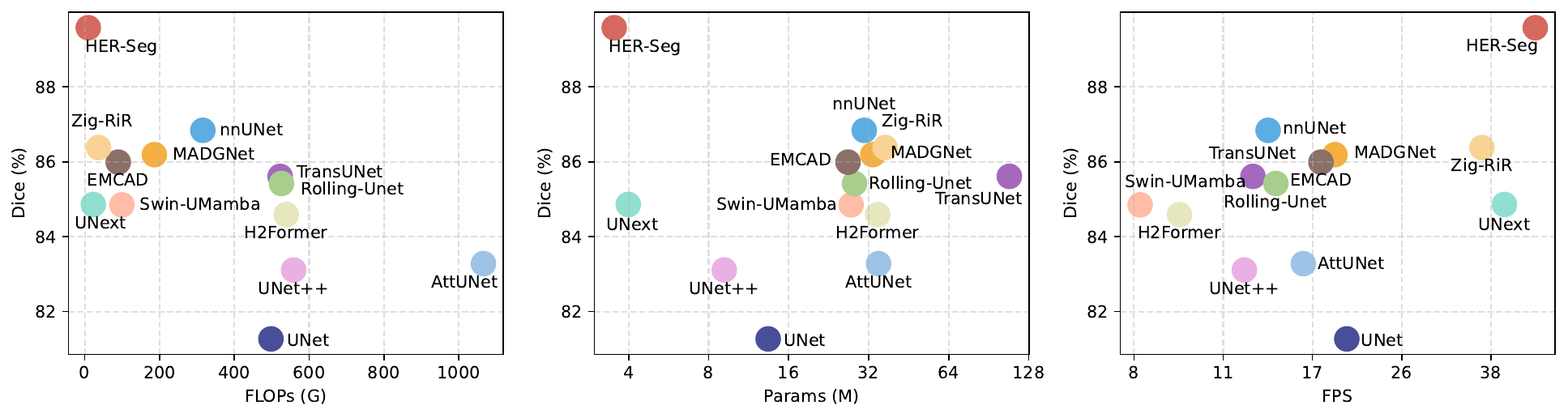}
  \caption{Comparisons of state-of-the-arts and our HER-Seg framework on computation cost. Compared with existing state-of-the-arts, our HER-Seg demonstrates superior performance with lower FLOPs and parameters, as well as faster inference speed.}
  \label{fig:intro_comp}
\end{figure*}

Recently, vision transformer (ViT) \cite{dosovitskiy2021an} has become a promising alternative to CNN in various computer vision tasks. It adopts the self-attention mechanism for long-sequence modeling, which enables the model to capture global dependencies across the entire input sequence by computing attention weights between all pairs of tokens. This global modeling capability has led many ViT-based U-shape frameworks \cite{cao2022swin, he2023h2former, chen2024transunet} to reveal superior performance in different medical image segmentation tasks compared to their CNN counterparts. However, the quadratic complexity $O(n^2)$ of the self-attention mechanism with respect to sequence length poses significant computational challenges for high-resolution medical image processing. As the input resolution increases, the computational cost grows exponentially, making it impractical for processing large-scale medical images. The dense attention computation requires calculating attention weights for all token pairs, which results in substantial computational overhead when dealing with high-resolution inputs such as $1024 \times 1024$ or larger medical images. These computational limitations significantly hinder the deployment of ViT-based segmentation models in clinical environments, particularly in mobile devices and edge computing scenarios where computation efficiency is crucial for real-time applications.

Furthermore, the decoder of current UNet-based methods \cite{ zhu2024selfreg, xu2024lb, chen2025zig} leverages a hierarchical bottom-up structure to progressively combine high-level and low-level semantic information through skip connections, making it suitable for mask generation with any size. This hierarchical decoding paradigm enables the model to recover fine-grained spatial details by gradually upsampling from coarse to fine resolutions across multiple scales. Additionally, \cite{zhou2019unet++} applied multiscale feature fusion to each skip connection layer, which enhances the precision of mask predictions by incorporating features from different resolution levels. However, when faced with high-resolution segmentation mask predictions, these hierarchical pyramid decoding operations require maintaining and processing large-size feature maps at multiple resolution levels simultaneously. This hierarchical structure inherently demands substantial memory allocation to store intermediate spatial information across different scales. Therefore, the memory requirements scale dramatically with input resolution, leading to prohibitive memory cost during training and inference.

To address the aforementioned limitations, we propose a holistically efficient encoder-decoder architecture for high-resolution medical image segmentation, named HER-Seg. Specifically, we first introduce the computation-efficient image encoder (CE-Encoder) that utilizes the dual-gated linear attention (DLA) mechanism to perform cascaded token filtering, selectively retaining important tokens while ignoring irrelevant ones to enhance attention computation efficiency. This approach enables modeling of long-range dependencies with linear complexity while maintaining sufficient representations. Then, we devise the memory-efficient mask decoder (ME-Decoder) that leverages cross-scale segmentation decoding to eliminate the demand for the hierarchical structure, significantly reducing the memory cost of high-resolution segmentation mask predictions. As illustrated in Fig. \ref{fig:intro_memory} and \ref{fig:intro_comp}, our HER-Seg outperforms state-of-the-arts in various high-resolution medical segmentation tasks with remarkable efficiency. 

The contributions are summarized as follows:
\begin{itemize}

\item We propose a holistically efficient HER-Seg framework that provides an end-to-end solution for high-resolution medical image segmentation with remarkable versatility across diverse medical modalities while maintaining low computational and memory cost.

\item We propose a CE-Encoder that employs DLA to perform intelligent cascaded token filtering, capturing global dependencies from long-range sequences with linear computation complexity and sufficient expressive power.

\item We devise a ME-Decoder that leverages cross-scale segmentation decoding to refine image embeddings, eliminating hierarchical structures and significantly reducing memory cost of high-resolution segmentation predictions.

\item We conduct extensive experiments on diverse high-resolution medical datasets, proving that our HER-Seg outperforms state-of-the-arts with only 0.59GB GPU memory and 9.39G FLOPs usage per $1024 \times 1024$ image.

\end{itemize}

\section{Related Work}

\begin{figure*}[!t]
  \centering
  \includegraphics[width=1\linewidth]{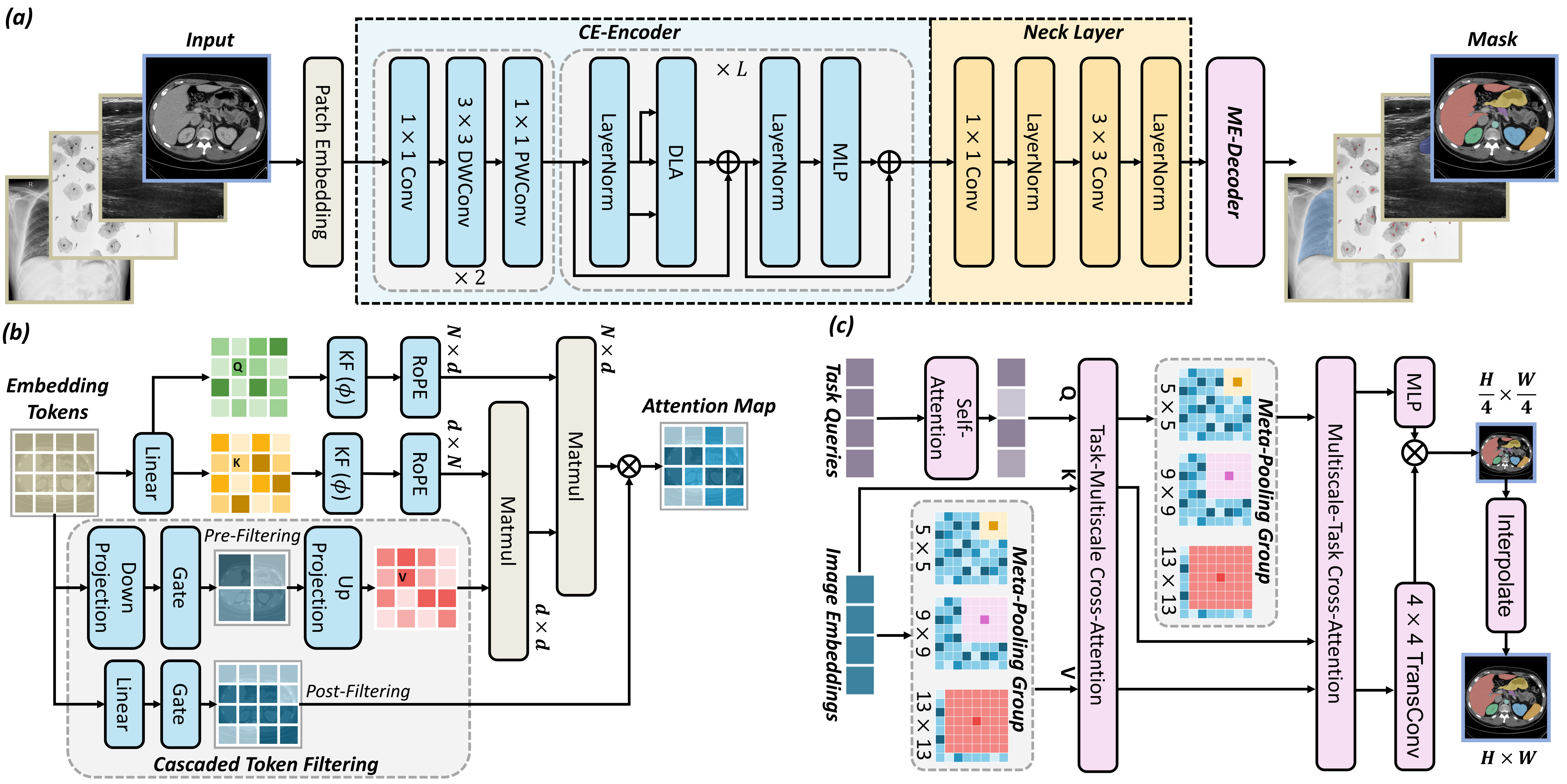}
  \caption{The overview of our HER-Seg framework for high-resolution medical segmentation. For ease of understanding, we show the case in multi-organ segmentation. HER-Seg leverages the cascaded token filtering and cross-scale decoding to achieve efficient computation and memory cost.}
  \label{fig:archi}
  \vspace{-1.0em}
\end{figure*}

\subsection{High-resolution Medical Image Segmentation}
High-resolution medical image segmentation has emerged as a critical technology for precise disease diagnosis and treatment planning. The ability to accurately segment high-resolution medical images is essential for detecting subtle lesions, analyzing tissue microstructures, and providing detailed anatomical information that is crucial for clinical decision-making. The encoder-decoder architecture, UNet \cite{ronneberger2015u}, has become a common fundamental design for many medical image segmentation models \cite{ ibtehaz2023acc, zhu2024selfreg}. To improve feature representations of the network, existing studies \cite{ schlemper2019attention, nam2024modality} usually utilized attention mechanisms to highlight salient features. Particularly, EMCAD \cite{rahman2024emcad} employed grouped channel and spatial gated attention mechanisms to efficiently catch intricate spatial relationships. However, such CNN-based frameworks suffered from a limited receptive field due to the kernel size, resulting in poor accuracy in high-resolution segmentation.

With superior ability to model long-range dependencies, ViT \cite{dosovitskiy2021an} surpassed CNN in various visual tasks. For medical image segmentation, Cao \textit{et al.} \cite{cao2022swin} proposed a pure transformer-based encoder-decoder architecture that used the hierarchical Swin Transformer with shifted windows for feature extraction and segmentation decoding operations. He \textit{et al.} \cite{he2023h2former} introduced a hyper-transformer block that combined multi-scale channel attention with self-attention to grasp local and global dependencies in medical images. The updated TransUNet \cite{chen2024transunet} constructed a coarse-to-fine transformer decoder to deal with small targets like tumors. Moreover, the large model capacity of ViT enhanced its generalization capabilities to unseen domains. The recent ViT-based SAM \cite{kirillov2023segment} has proved its outstanding zero-shot learning ability in diverse medical domains \cite{ma2024segment}. Despite their advantages, the rapid increase in the number of parameters in ViT caused huge computational cost during training and deployment, which limited its applications in high-resolution medical segmentation.

\subsection{Efficient Medical Image Segmentation}
With the increasing deployment of medical image analysis systems in resource-constrained environments, the development of lightweight medical segmentation approaches has become crucial for democratizing access to advanced diagnostic tools and enabling real-time medical image analysis. To this end, directly downscaling the embedding dimensions and network layers was a common approach to decrease the model size \cite{xiong2024efficientsam}. In addition, Mamba \cite{gu2023mamba} integrated state space model, an advanced recurrent neural network, to model long-range dependencies with linear complexity. The recent works have demonstrated the effectiveness of Mamba in diverse high-resolution medical image segmentation tasks \cite{xing2024segmamba, liu2024swin}.

It is noteworthy that most existing lightweight medical segmentation approaches primarily focus on optimizing the encoder components while neglecting the efficiency of the decoder parts. Current studies utilized depthwise convolutions \cite{hao2024emf, chen2024tinyu}, advanced multi-layer perceptrons \cite{valanarasu2022unext, liu2024rolling} and Kolmogorov-Arnold networks \cite{li2024u} to reduce computation cost in decoding layers of U-shape architectures. However, for high-resolution medical image segmentation, such multi-level decoding processes significantly increased memory costs. These hierarchical decoding operations require maintaining and processing large-size feature maps at multiple resolution levels simultaneously. Different from these methods, our proposed HER-Seg framework adopts a holistically efficient approach that simultaneously optimizes both encoder and decoder components to achieve comprehensive efficiency improvements in high-resolution medical image segmentation.

\section{Methodology}

\subsection{Overview of HER-Seg}

We present the HER-Seg framework to provide holistically efficient high-resolution segmentation with superior performance across diverse medical imaging modalities, as illustrated in Fig. \ref{fig:archi}(a). Our goal is to train an end-to-end model $f_\theta: x \rightarrow y$, where $\theta$ represents learned parameters, $x$ is the given image and $y$ is the predicted segmentation mask. Each pixel in this mask is assigned to a hard label based on the predefined class list. Our HER-Seg is built upon the encoder-decoder architecture, integrating a CE-Encoder that utilizes dual-gated linear attention mechanism to perform cascaded token filtering, selectively retaining important tokens while ignoring irrelevant ones to enhance attention computation efficiency. Moreover, we leverage a ME-Decoder to eliminate the demand for a hierarchical structure, significantly reducing memory cost during large-size segmentation decoding, making the framework practical for real-world clinical deployment. Finally, we utilize feature distillation during pretraining to reduce training time and fully unleash HER-Seg's potential.

\subsection{Computation-Efficient Image Encoder}

Existing efficient ViTs adopt group attention \cite{liu2023efficientvit}, depthwise convolution \cite{wang2024repvit}, or multi-scale linear attention \cite{cai2023efficientvit} to decrease the computational overhead of attention layers. However, these methods primarily focus on architectural modifications while neglecting the efficiency of input token processing. Inspired by the sparse saliency nature of medical images, where only certain regions contain diagnostically relevant information, we propose a CE-Encoder for efficient high-resolution feature representations. As shown in Fig. \ref{fig:archi}(b), the CE-Encoder is designed with a dual-gated linear attention (DLA) mechanism that performs intelligent token pre-screening, where the key innovation lies in selectively retaining important tokens while filtering out irrelevant ones before attention computation. This approach enables efficient modeling of long-range dependencies with linear complexity while maintaining sufficient expressive power for high-resolution image segmentation.

To build up our CE-Encoder, we first employ two channel-interactive blocks to learn low-level representation efficiently. Specifically, each block includes a $1 \times 1$ convolution $\mathcal{F}_{\rm Conv}^{\rm 1 \times 1}$ for channel expansion. A $3 \times 3$ depthwise convolution $\mathcal{F}_{\rm DWConv}^{\rm 3 \times 3}$ is followed by a $1 \times 1$ pointwise convolution $\mathcal{F}_{\rm PWConv}^{\rm 1 \times 1}$ for channel communication. Given a set of input patch embeddings $x \in \mathbb{R}^{\frac{H}{S} \times \frac{W}{S} \times  C}$, where $H$ and $W$ represent the height and width of the input image, $S$ is the predefined patch size and $C$ is channel. The computation can be defined by:
\begin{equation}
    x \gets \sigma(\mathcal{F}_{\rm PWConv}^{\rm 1 \times 1}(\sigma(\mathcal{F}_{\rm DWConv}^{\rm 3 \times 3}(\sigma(\mathcal{F}_{\rm Conv}^{\rm 1 \times 1}(x))))) + x),
\end{equation}
where $\sigma$ stands for an activation function (\textit{e.g.,} GELU). This design enables the channel-interactive blocks to enhance local feature extraction capabilities by introducing inductive bias of spatial structural information, which is particularly beneficial for medical images where local texture patterns and spatial relationships are crucial for accurate segmentation. The lightweight nature of these blocks ensures computation efficiency while maintaining sufficient representational power for subsequent attention processing.

After extracting local structural features through channel-interactive blocks, we proceed to capture global dependencies for comprehensive feature representation. The standard self-attention mechanism is defined by:
\begin{equation}\label{eq2}
    \begin{aligned}
       V'&=\mathrm{softmax}(\frac{Q K^{\top}}{\sqrt{d}})V,\\
        Q&=xW_Q, K=xW_K, V=xW_V,  
    \end{aligned}
\end{equation}
where $W_Q, W_K, W_V \in \mathbb{R}^{C\times d}$ are the learnable linear projection matrices with $d$ representing the projection dimension. We observe that self-attention computes the similarities between all query-key pairs to update input $x$. This operation can be achieved through the gating mechanism. Therefore, our DLA reformulates Equation \ref{eq2} as:
\begin{equation}
    V_{i}'=\sum_{j=1}^{N} \frac{\phi(Q_{i}) \phi(K_{j})^{\top}}{\sum_{j=1}^{N} \phi(Q_{i}) \phi(K_{j})^{\top}} V_{j}^{\rm filtered}, 
\end{equation}
where $V_i$ represents the $i$-th row of matrix $V$, $\phi$ is the kernel function following \cite{katharopoulos2020transformers} and $N=H \times W/S^2$. Moreover, we enhance the query and key representations by incorporating rotary position embedding (RoPE) \cite{su2024roformer}. This rotational encoding preserves the relative distance between tokens while maintaining translation invariance. $V^{\rm filtered}$ is obtained by a pre-filtering network:
\begin{equation}
    V^{\rm filtered}= f(VW_{\rm proj}^{\rm down})W_{\rm proj}^{\rm up},
\end{equation}
where $f(\cdot)$ is the gate function (e.g., SiLU). The pre-filtering network leverages nonlinear transformations to suppress irrelevant tokens, enhancing the efficiency of value representation. Then, we apply a post-filtering network to dynamically discern the utility of input $x$, further filtering redundant information. The final image embedding $h$ can be calculated by:
\begin{equation}
    h = \mathcal{F}_{\rm MLP}(\mathcal{F}_{\rm LN}(f(xW_x)V')) + x,
\end{equation}
where $h \in \mathbb{R}^{N \times C}$, $W_x \in \mathbb{R}^{C \times d}$, $\mathcal{F}_{\rm MLP}(\cdot)$ is the multi-layer perceptron and $\mathcal{F}_{\rm LN}(\cdot)$ is LayerNorm. Overall, the proposed CE-Encoder adopts channel-interactive blocks to efficiently learn low-level representations and DLA to perform cascaded token filtering, capturing long-range visual dependencies with linear complexity, enabling scalable and efficient attention computation for high-resolution medical image processing.

\subsection{Memory-Efficient Mask Decoder}
Current efficient medical segmentation approaches still retain the classical U-shape architecture with hierarchical bottom-up decoding operations to progressively combine multi-level semantic information for mask generation. However, when processing high-resolution medical images, these hierarchical decoding paradigms require maintaining and processing large-size feature maps at multiple resolution levels simultaneously, leading to substantial memory allocation demands that scale dramatically with input resolution. To address this critical limitation, we devise a ME-Decoder that leverages cross-scale segmentation decoding to eliminate the demand for hierarchical pyramid structures while maintaining superior segmentation performance, as shown in Fig. \ref{fig:archi}(c). Unlike conventional decoders that process features across multiple resolution levels, our ME-Decoder performs all decoding operations in high-dimensional but spatially compact image embeddings, significantly reducing memory footprint during high-resolution segmentation mask predictions. We first create task query embeddings $q \in \mathbb{R}^{c \times 256}$ to learn the decoding information, where $c$ is the number of prediction categories. These query embeddings serve as compact representations that encode class-specific segmentation knowledge without requiring large spatial dimensions. Then, we apply self-attention to update the task query $q$, enabling the queries to refine their representational capacity through internal interactions. Following query refinement, we employ bidirectional multiscale cross-attention layers between the updated queries $q$ and image embeddings $h$ to perform cross-scale information exchange. This bidirectional interaction can be defined by:
\begin{equation}
    q \gets \mathrm{softmax}(\frac{q (h+\psi)^{\top}}{\sqrt{d}})\delta(h)+q,
\end{equation}
\begin{equation}
    h \gets \mathrm{softmax}(\frac{(\delta(h)+\psi) (q)^{\top}}{\sqrt{d}})q + \delta(h),
\end{equation}
where $\psi$ denotes positional encoding that enhances the geometric location awareness, and $\delta(\cdot)$ represents a meta-pooling group comprising three parallel meta-pooling operators \cite{yu2022metaformer} with different kernel sizes to capture semantic information at diverse scales without introducing additional parameters. This cross-scale mechanism enables the decoder to aggregate multiscale contextual information within compact embeddings rather than across multiple resolution levels. The final segmentation mask $y$ generation is achieved through:
\begin{equation} 
y= \mathrm{softmax}(\mathcal{F}_{\rm inter}(\mathcal{F}_{\rm trans}(h) \cdot \mathcal{F}_{\rm MLP}(q))),
\end{equation}
where $\mathcal{F}_{\rm trans}(\cdot)$ stands for a $4 \times 4$ transpose convolution and $\mathcal{F}_{\rm inter}(\cdot)$ is a bilinear interpolation function that directly restores the mask to the original input resolution. By performing all decoding operations in high-dimensional but spatially compact embeddings and eliminating the need for hierarchical pyramid structures, our ME-Decoder substantially reduces memory consumption during high-resolution segmentation mask predictions. This design enables efficient processing of large-scale medical images while maintaining the model's capability to capture fine-grained spatial details.

\begin{table*}[!t]
    \centering
    \small
    \caption{Comparison with state-of-the-arts on 2D medical image segmentation.}
    {\scalebox{0.93}{
    \begin{tabular}{l|ccc|cccccccccc}
    \hline
     \multirow{2}{*}{Methods} & \multirow{2}{*}{\#Mem} & \multirow{2}{*}{\#FLOPs} & \multirow{2}{*}{\#Params} & \multicolumn{2}{c}{ISIC-2018} & \multicolumn{2}{c}{PCXA} & \multicolumn{2}{c}{REFUGE} & \multicolumn{2}{c}{UDIAT} & \multicolumn{2}{c}{DSB-2018} \\
    \cline{5-14}
    & & & & Dice & mIoU & Dice & mIoU &  Dice & mIoU & Dice & mIoU & Dice & mIoU\\
    \hline
    U-Net \cite{ronneberger2015u} & 7.63GB & 497.91G & 13.40M & 80.36 & 70.23 & 96.30 & 92.94 & 86.86 & 77.72 & 69.66 & 57.61 & 87.52 & 80.39\\ 
    U-Net++ \cite{zhou2019unet++} & 9.82GB & 558.46G & 9.16M & 82.06 & 72.95 & 96.43 & 93.18 & 88.49 & 79.92 & 70.66 & 58.83 & 88.97 & 82.67\\ 
    AttUNet \cite{schlemper2019attention} & 12.18GB & 1066.11G & 34.87M & 82.40 & 73.42 & 96.19 & 92.74 & 89.04 & 80.67 & 72.06 & 60.12 & 89.03 & 82.81\\
    nnUNet \cite{isensee2021nnu} & 7.15GB & 315.70G & 30.80M & 85.08 & 76.74 & 96.74 & 93.78 & 89.13 & 80.76 & \underline{77.07} & \underline{70.08} & 90.48 & 82.89\\
    UNext \cite{valanarasu2022unext} & \underline{1.45GB} & \underline{22.75G} & \underline{3.99M} & 85.27 & 77.36 & 96.42 & 93.27 & 86.14 & 76.31 & 76.45 & 66.82 & 90.16 & 82.63\\ 
    TransUNet \cite{chen2024transunet} & 15.69GB & 522.45G & 108.27M & 86.44 & 78.89 & 96.56 & 93.42 & 88.43 & 79.80 & 72.18 & 60.85 & 91.12 & 84.31\\
    H2Former \cite{he2023h2former} & 37.37GB & 538.58G & 34.60M & 85.20 & 77.33 & 96.45 & 93.23 & 89.22 & 81.09 & 72.02 & 60.05 & 90.79 & 83.86\\
    Swin-UMamba \cite{liu2024swin} & 2.18GB & 99.35G & 27.49M & 85.34 & 76.64 & 96.72 & 93.74 & 87.93 & 78.69 & 70.72 & 58.97 & 90.04 & 82.59\\
    Rolling-Unet \cite{liu2024swin} & 8.30GB & 525.93G & 28.32M & 85.13 & 77.03 & 96.63 & 93.51 & 88.19 & 79.39 & 73.24 & 62.06 & 90.93 & 83.98\\
    MADGNet \cite{nam2024modality} & 3.18GB & 186.11G & 33.15M & 85.96 & 77.95 & 96.55 & 93.40 & 88.65 & 80.09 & 73.91 & 62.62 & \underline{91.20} & \underline{84.49}\\
    EMCAD \cite{rahman2024emcad} & 4.03GB & 89.53G & 26.76M & \underline{86.97} & \underline{79.52} & 96.74 & 93.77 & \underline{89.44} & \underline{81.31} & 73.13 & 61.74 & 90.62 & 83.76\\
    Zig-RiR \cite{chen2025zig} & 2.42GB & 50.27G & 24.58M & 86.52 & 78.71 & \underline{96.78} & \underline{93.82} & 88.97 & 80.45 & 74.28 & 63.15 & 91.05 & 84.17\\
    \hline
    HER-Seg & \textbf{0.59GB} & \textbf{9.39G} & \textbf{3.67M} & \textbf{88.63} & \textbf{81.62} & \textbf{96.86} & \textbf{94.00} & \textbf{89.81} & \textbf{81.84} & \textbf{85.50} & \textbf{77.30} & \textbf{92.10} & \textbf{85.78}\\ 
    \hline
    \end{tabular}}}
    \label{tab1}
\end{table*}

\begin{figure*}[!t]
  \centering
  \includegraphics[width=0.9\linewidth]{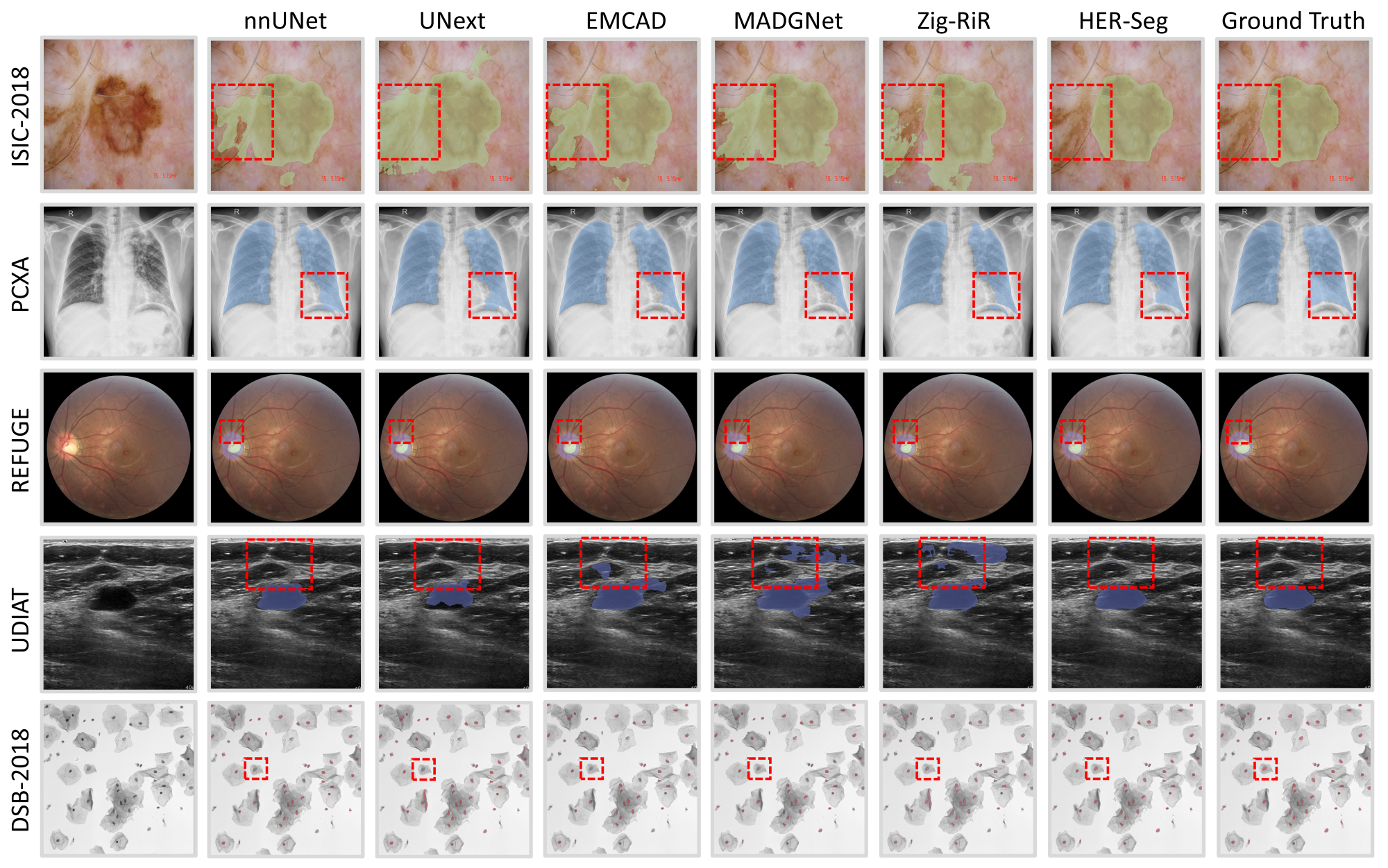}
  \caption{Visualization of high-reslution 2D medical segmentation. Our HER-Seg exhibits the best results, recognizing more lesion areas and cells with accurate categories and boundaries while having fewer false positives.}
  \label{fig:t1}
\end{figure*}

\subsection{Architecture Optimization}
To construct our HER-Seg framework, we first adopt the CE-Encoder to achieve efficient feature extraction from high-resolution medical images. It begins with patch embedding to tokenize the input medical image into sequential patches, followed by 2 channel-interactive layers and $L$ transformer blocks. The expansion ratio in MLP is generally set to 4 in the original ViT, which makes the hidden dimension $4\times$ wider than the embedding dimension $C$. Despite this setting improving model capacity, the wider dimension consumes a significant portion of parameters and memory resources during high-resolution image processing. To mitigate this bottleneck, our HER-Seg sets the expansion ratio to 2 in both DWConv and MLP layers. Moreover, deeper models are more likely to lead to overfitting on medical datasets as they usually contain limited fully-annotated labels due to the expensive pixel-level annotations. Therefore, our CE-Encoder adopts a narrower embedding dimension and shallow structure: $C=96, L=10$ to achieve optimal efficiency-performance trade-off.

Furthermore, we adopt a simple but efficient neck layer that consists of a $3\times3$ convolution followed by a $1\times1$ convolution to expand the dimension of the image embedding to $D$, further enhancing the model capacity and expressive power. The refined feature map is then fed into the ME-Decoder for final segmentation mask generation. The predicted segmentation mask $y$ is supervised by the combination
of cross-entropy loss $\mathcal{L}_{\rm CE}$ and dice loss $\mathcal{L}_{\rm Dice}$, as follows:
\begin{equation}
    \mathcal{L}_{\rm Seg} = \lambda_1\mathcal{L}_{\rm CE} + \lambda_2\mathcal{L}_{\rm Dice},
\end{equation}
where $\lambda_1$ and $\lambda_2$ are the coefficients to balance these two loss terms. In this way, the holistically efficient design enables HER-Seg to handle high-resolution medical images with low computation and memory cost while maintaining competitive segmentation performance across diverse medical modalities.

\section{Experiments}
\subsection{Datasets and Implementations}
\subsubsection{Datasets}
To validate the effectiveness of the proposed Co-Seg++, we conduct comprehensive evaluations across 7 different high-resolution medical modalities, as follows:

\noindent \textbf{ISIC-2018} \cite{tschandl2018ham10000, codella2019skin} is designed to aid in the development of automated systems for the skin melanoma diagnosis. It contains 3694 high-quality dermoscopic images with the highest resolution of $2304 \times 3072$.

\noindent \textbf{PCXA} \cite{jaeger2013automatic, candemir2013lung} is a lung segmentation dataset for automatic tuberculosis screening, including 704 chest x-rays with the resolution of $4020 \times 4892$.

\noindent \textbf{REFUGE} \cite{orlando2020refuge} dataset, derived from the Retinal Fundus Glaucoma Challenge, aimed at advancing automated glaucoma assessment. It consists of 1200 retinal fundus images, with the resolution of $2124 \times 2056$, for optic disc and cup segmentation.

\noindent \textbf{UDIAT} \cite{yap2017automated} dataset includes 163 ultrasound images with the resolution of $760 \times 570$. We use it to evaluate models in the automatic breast lesion segmentation task.

\noindent \textbf{DSB-2018} \cite{caicedo2019nucleus} is a part of the 2018 Data Science Bowl challenge, including 670 microscopic slides collected from different nuclei types, staining protocols and image conditions, with the highest resolution of $1024 \times 1024$.

\noindent \textbf{Synapse} \cite{Synapse2015xu} is a multi-organ segmentation dataset. It collects 3,779 axial contrast-enhanced slices from 30 abdominal CT scans with the resolution of $512 \times 512$. Based on the configuration of many studies \cite{rahman2024emcad, chen2024transunet}, we also segment the same eight abdominal organs in our experiments.

\noindent \textbf{CVC-ClinicDB} \cite{bernal2015wm, vazquez2017benchmark} and \textbf{CVC-ColonDB} \cite{tajbakhsh2015automated} are colonoscopy video datasets. They contain 612 and 380 polyp frames with the resolution of $884 \times 1280$.

\subsubsection{Implementation Details}

We perform all experiments on 1 NVIDIA A6000 GPU with PyTorch. We adopt the optimizer using Adam with a learning rate of $1\times10^{-4}$ and apply the exponential decay strategy to adjust the learning rate, where the factor is set as 0.98. The batch size and epochs are set to 16 and 200, respectively. We evaluate all baselines and our HER-Seg on the high resolution of $1024 \times 1024$. The expanded dimension $D$ of the neck layer is set as 256 based on \cite{ravi2024sam}. The loss coefficients $\lambda_1$ and $\lambda_2$ are set as 1. We adopt common train-val-test splits \cite{rahman2024emcad, nam2024modality, chen2025zig} for all datasets. For 3D segmentation tasks, we follow previous studies \cite{chen2024transunet, rahman2024emcad} that convert 3D volumes to 2D slices. This conversion protocol ensures the same memory requirements between 2D and 3D applications while maintaining the computation and memory efficiency of our HER-Seg framework.

\subsubsection{Evaluation Metrics}
To perform a comprehensive evaluation of medical segmentation, we apply different metrics in terms of 2D, 3D and video segmentation tasks. For 2D image and video segmentation, we select two standard metrics: Dice coefficient and mean intersection over union (mIoU). For 3D volume segmentation, we additionally compute the Hausdorff distance (HD). Except for HD, which measures the distance between predicted and ground truth boundary points, higher scores for these metrics indicate better segmentation quality.

\subsection{Comparison on 2D Medical Image Segmentation}
To comprehensively evaluate the performance of HER-Seg, we conduct extensive comparisons with state-of-the-art methods on 2D medical image segmentation across five diverse datasets. As shown in Table \ref{tab1}, traditional encoder-decoder architectures \cite{ronneberger2015u, zhou2019unet++} generally exhibit limited performance due to their inefficient memory utilization and computational overhead. Recent efficient architectures such as UNext \cite{valanarasu2022unext} and Swin-UMamba \cite{liu2024swin} achieve better efficiency-performance trade-offs. Remarkably, HER-Seg substantially outperforms all baseline methods, with a P-value $<$ 0.005 while maintaining the lowest computation cost. For the challenging ultrasound segmentation, HER-Seg reveals exceptional performance with 85.50\% Dice and 77.30\% mIoU, representing a substantial 8.43\% Dice improvement over the best baseline nnUNet \cite{isensee2021nnu}.

\begin{table*}[!t]
    \centering
    \small
    \caption{Comparison with state-of-the-arts on 3D medical image segmentation.}
    {\scalebox{0.96}{
    \begin{tabular}{l|ccc|ccccccccc}
    \hline
     \multirow{2}{*}{Methods} & \multicolumn{3}{c|}{Average} & \multirow{2}{*}{Aorta} & \multirow{2}{*}{Gallbladder} & \multirow{2}{*}{Kidney (L)} & \multirow{2}{*}{Liver} 
     & \multirow{2}{*}{Pancreas} & \multirow{2}{*}{Spleen} & \multirow{2}{*}{Stomach} & \multirow{2}{*}{Kidney (R)} \\
    \cline{2-4}
    & Dice & mIoU & HD &   &  &  &  &   &  &    \\
    \hline
    U-Net \cite{ronneberger2015u} & 74.17 & 68.06 & 98.39 & 86.24 & 58.79 & 81.83 & 88.96 & 49.31 & 81.57 & 66.50 & 80.17 \\
    U-Net++ \cite{zhou2019unet++} & 78.11 & 71.25 & 64.44 & 91.82 & 66.57 & 88.33 & 92.42 & 50.08 & 85.13 & 65.91 & 84.65 \\ 
    AttUNet \cite{schlemper2019attention} & 78.16 & 71.21 & 63.78 & 91.41 & 60.52 & 85.61 & 91.81 & 53.06 & 86.71 & 70.44 & 85.71 \\
    nnUNet \cite{isensee2021nnu} & 81.48 & 74.79 & 49.57 & \underline{92.33} & 65.08 & 87.52 & 92.55 & 60.28 & 90.89 & 77.35 & \underline{85.87}\\
    UNext \cite{valanarasu2022unext} & 79.27 & 72.03 & 51.57 & 90.38 & 65.66 & 87.25 & 90.83 & 58.54 & 87.47 & 70.49 & 83.53 \\
    TransUNet \cite{chen2024transunet} & 81.31 & 74.15 & 53.22 & 91.37 & 67.84 & 88.26 & 89.79 & 64.05 & 89.51 & 76.13 & 83.55 \\
    H2Former \cite{he2023h2former} & 78.32 & 70.83 & 52.04 & 90.80 & 64.19 & 87.94 & 90.91 & 51.93 & 87.63 & 70.60 & 82.59\\
    Swin-UMamba \cite{liu2024swin} & 76.71 & 69.14 & 54.66 & 90.37 & 60.76 & 87.22 & 89.55 & 51.10 & 85.85 & 67.41 & 81.44\\
    Rolling-Unet \cite{liu2024rolling}  & 80.75 & 73.61 & 50.32 & 90.63 & 67.25 & 86.98 & 90.24 & 60.16 & 89.82 & 77.32 & 83.56
    \\
    MADGNet \cite{nam2024modality} & 81.37 & 74.43 & 50.76 & 91.87 & 68.19 & 87.34 & 91.31 & 59.58 & \underline{91.38} & 75.88 & 85.40 \\
    EMCAD \cite{rahman2024emcad} & \underline{83.24} & \underline{76.45} & \underline{43.31} & 91.95 & 69.59 & \underline{90.79} & \underline{92.60} & 65.47 & \textbf{91.53} & \textbf{79.95} & 84.03\\
    Zig-RiR \cite{chen2025zig} & 82.47 & 75.68 & 46.15 & 92.18 & \underline{70.95} & 89.51 & 92.19 & \underline{67.82} & 90.15 & 77.86 & 83.38\\
    \hline
    HER-Seg & \textbf{84.25} & \textbf{77.57} & \textbf{36.09} & \textbf{92.58} & \textbf{72.33} & \textbf{91.46} & \textbf{93.75} & \textbf{70.06} & 89.08 & \underline{78.59} & \textbf{86.17}\\ 
    \hline
    \end{tabular}}}
    \label{tab2}
\end{table*}

\begin{table}[!t]
    \centering
    \small
    \setlength\tabcolsep{9pt}
    \caption{Comparison on medical video segmentation.}
    {\scalebox{1}{
    \begin{tabular}{l|cccc}
    \hline
    \multirow{2}{*}{Methods} & \multicolumn{2}{c}{CVC-ColonDB} & \multicolumn{2}{c}{CVC-ClinicDB} \\
    \cline{2-5}
     & Dice & mIoU & Dice & mIoU\\
    \hline
    U-Net \cite{ronneberger2015u} & 69.27 & 61.84 & 86.02 & 78.07\\  
    U-Net++ \cite{zhou2019unet++} & 71.14 & 62.03 & 88.98 & 82.24\\ 
    AttUNet \cite{schlemper2019attention} & 72.50 & 63.20 & 86.85 & 78.50 \\
    nnUNet \cite{isensee2021nnu} & \underline{83.67} & \underline{76.28} & 91.06 & 85.02\\  
    UNext \cite{valanarasu2022unext} & 76.63 & 69.27 & 88.52 & 81.46\\  
    TransUNet \cite{chen2024transunet} & 79.54 & 71.06 & 89.27 & 82.98\\
    H2Former \cite{he2023h2former} & 76.08 & 65.89 & 88.63 & 81.71\\
    Swin-UMamba \cite{liu2024swin} & 80.61 & 72.72 & 90.73 & 84.81\\
    Rolling-Unet \cite{liu2024rolling} & 78.29 & 70.25 & 90.21 & 83.97\\ 
    MADGNet \cite{nam2024modality} & 80.79 & 72.93 & 91.10 & 85.08 \\
    EMCAD \cite{rahman2024emcad} & 76.65 & 69.41 & \underline{91.16} & \underline{85.55}\\
    Zig-RiR \cite{chen2025zig} & 81.24 & 73.55 & 90.87 & 84.63\\
    \hline
    HER-Seg & \textbf{85.03} & \textbf{78.21} & \textbf{94.42} & \textbf{89.79}\\ 
    \hline
    \end{tabular}}}
    \label{tab3}
\end{table}

Moreover, HER-Seg demonstrates remarkable efficiency advantages, utilizing only 0.59GB GPU memory per batch during fine-tuning on $1024 \times 1024$ high-resolution medical images, which is 2.46$\times$ more memory-efficient than the second-best UNext \cite{valanarasu2022unext} and 6.8$\times$ more efficient than the high-performance EMCAD \cite{rahman2024emcad}. In terms of computational complexity, HER-Seg requires merely 9.39G FLOPs, representing a 5.35× reduction compared to the efficient Zig-RiR \cite{chen2025zig} and 53.0$\times$ reduction compared to the traditional U-Net \cite{ronneberger2015u}. With only 3.67M parameters, HER-Seg achieves a 1.09$\times$ parameter reduction compared to the compact UNext while maintaining superior performance. These comprehensive comparisons validate the holistic efficiency and superior performance of HER-Seg for high-resolution medical image segmentation.

\subsection{Comparison on 3D Medical Image Segmentation}

To further validate the effectiveness of HER-Seg on volumetric medical data, we conduct comprehensive evaluations on the Synapse dataset with eight abdominal organs. As shown in Table \ref{tab2}, the recent state-of-the-art method EMCAD \cite{rahman2024emcad} demonstrates competitive results with 83.24\% average Dice. In contrast, our HER-Seg consistently outperforms all baseline methods across multiple evaluation metrics, achieving superior average performance with a P-value $<$ 0.005. Specfically, HER-Seg achieves 70.06\% Dice, representing a substantial 2.24\% improvement over the second-best Zig-RiR \cite{chen2025zig} (67.82\%) and 4.59\% improvement over EMCAD \cite{rahman2024emcad} (65.47\%).
In addition, HER-Seg demonstrates remarkable boundary precision with the lowest average HD of 36.09, indicating superior spatial accuracy compared to all competing methods. This represents a significant 7.22 improvement over EMCAD and 10.06 improvement over Zig-RiR, highlighting the effectiveness of our memory-efficient mask decoder in preserving fine-grained anatomical details. These results validate the effectiveness of HER-Seg for volumetric medical analysis.

\begin{figure}[!t]
  \centering
  \includegraphics[width=1\linewidth]{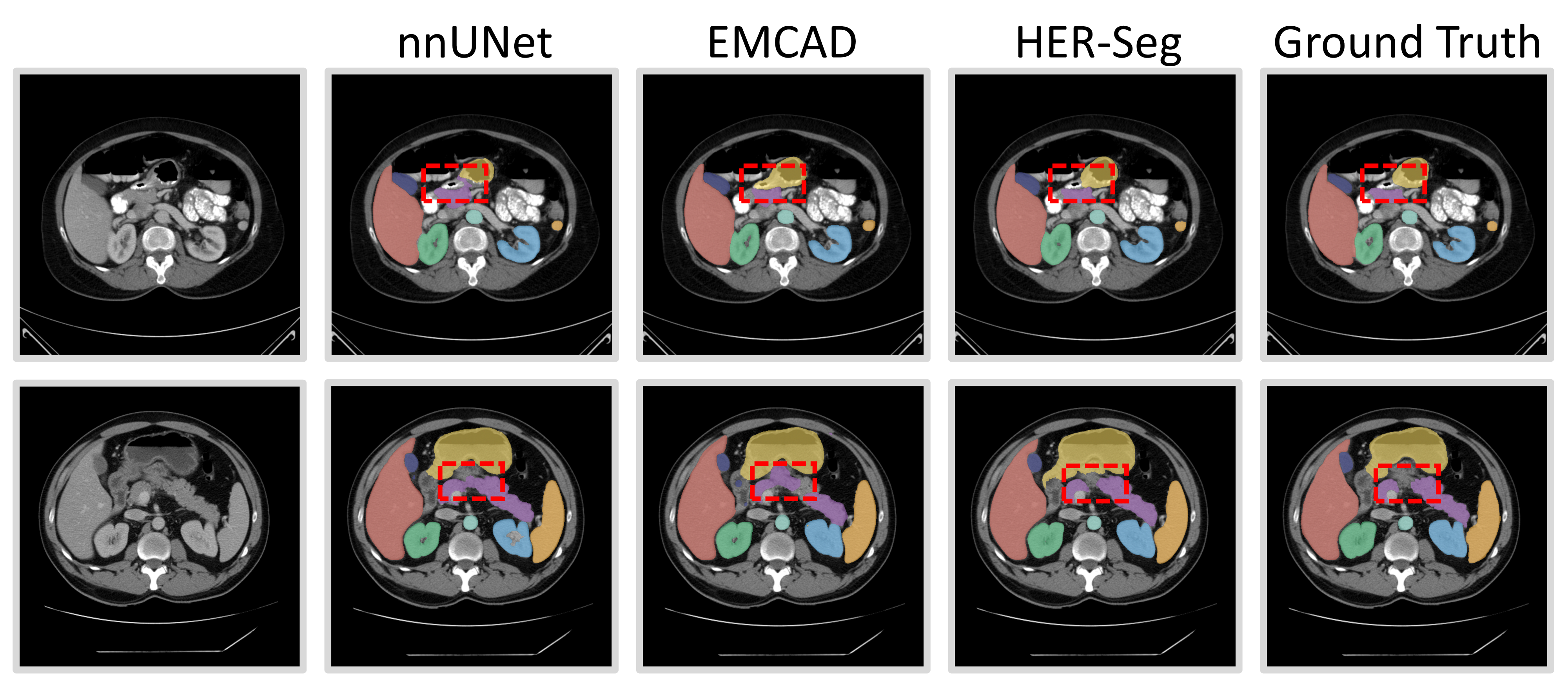}
  \caption{Visualization of high-resolution 3D multi-organ segmentation. Our HER-Seg performs better in identifying the boundary of each organ.}
  \label{fig:t2}
\end{figure}

\begin{figure}[!t]
  \centering
  \includegraphics[width=1\linewidth]{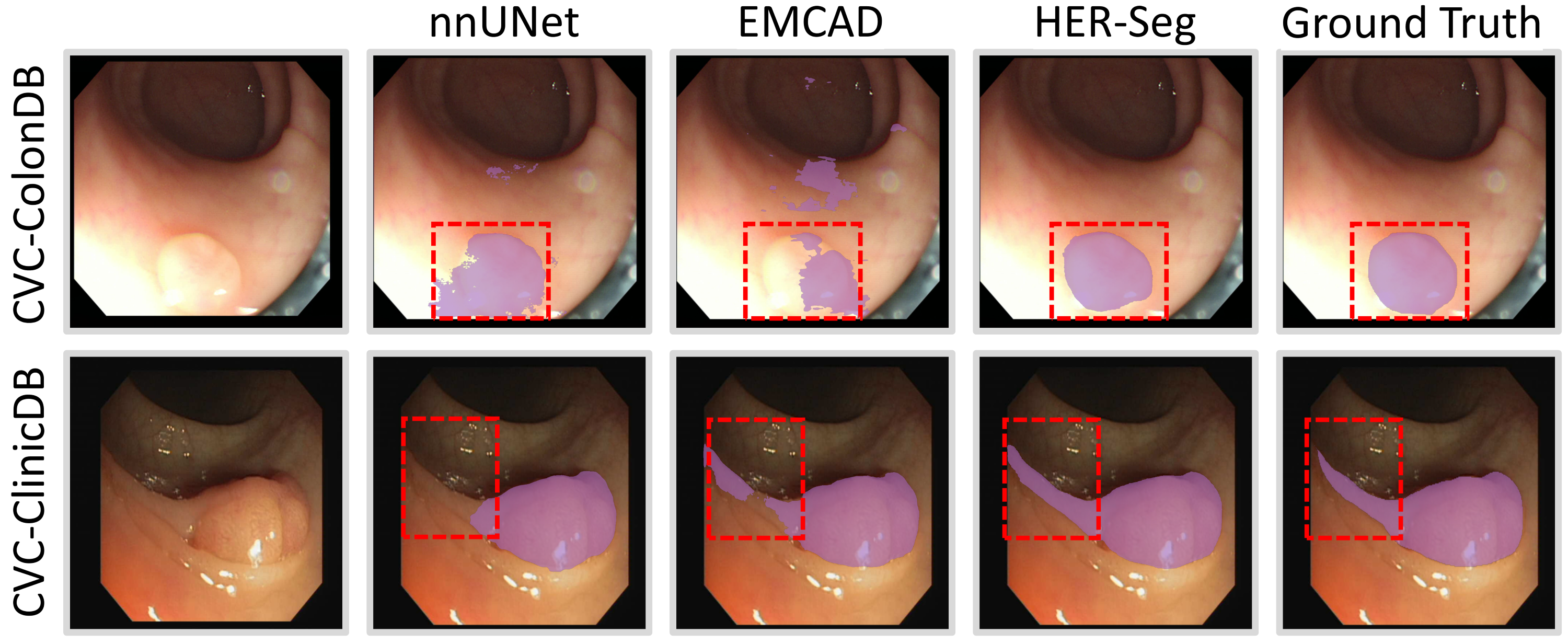}
  \caption{Visualization of high-resolution medical video segmentation. Our HER-Seg shows better performance in complex image conditions.}
  \label{fig:t3}
\end{figure}

\begin{table*}[!t]
    \centering
    \small
    \caption{Ablation study of our HER-Seg on high-resolution skin lesion, multi-organ, and polyp segmentation datasets.}
    {\scalebox{1}{
    \begin{tabular}{ccc|ccc|cc|ccc|cc}
    \hline
    \multirow{2}{*}{CE-Encoder} & \multirow{2}{*}{DLA} & \multirow{2}{*}{ME-Decoder} & \multirow{2}{*}{\#Mem} & \multirow{2}{*}{\#FLOPs} & \multirow{2}{*}{\#Params} & \multicolumn{2}{c|}{ISIC-2018} &  \multicolumn{3}{c|}{Synapse} & \multicolumn{2}{c}{CVC-ColonDB}\\
    \cline{7-13}
    &  &  &  &  &  & Dice & mIoU & Dice & mIoU & HD & Dice & mIoU\\
    \hline
     &  &  & 7.63GB & 497.71G & 13.84M & 77.12 & 70.23 & 74.17 & 68.06 & 98.39 & 69.27 & 61.84\\ 
    \checkmark &  &  & 5.85GB & 324.93G & 9.02M & 82.45 & 74.86 & 78.34 & 71.52 & 72.18 & 76.19 & 68.73\\ 
    \checkmark & \checkmark & & 2.57GB & 97.41G & 5.59M & 84.76 & 77.94 & 81.62 & 74.87 & 58.46 & 79.64 & 72.85\\
    \checkmark & \checkmark & \checkmark & 0.59GB & 9.39G & 3.67M & 88.63 & 81.62 & 84.25 & 77.57 &  36.09 & 85.03 & 78.21\\
    \hline
    \end{tabular}}}
    \label{tab:5}
\end{table*}

\subsection{Comparison on Medical Video Segmentation}

To evaluate the temporal consistency and robustness of HER-Seg, we conduct extensive experiments on medical video segmentation using two challenging polyp segmentation datasets: CVC-ColonDB and CVC-ClinicDB. As shown in Table \ref{tab3}, our HER-Seg framework demonstrates exceptional performance across both video datasets, significantly outperforming all baseline methods with a P-value $<$ 0.001. On the CVC-ColonDB dataset, HER-Seg achieves 85.03\% Dice and 78.21\% mIoU, representing a substantial 1.36\% Dice improvement over the second-best nnUNet \cite{isensee2021nnu} and 1.93\% improvement over Zig-RiR \cite{chen2025zig}. Notably, the method shows particularly strong performance in mIoU metrics, with 4.24\% improvement on CVC-ClinicDB compared to EMCAD, indicating superior region-wise segmentation accuracy. These comprehensive evaluations validate that HER-Seg's holistic efficiency design not only reduces computational overhead but also enhances segmentation quality in medical video scenarios, making it highly suitable for real-time clinical applications.

\begin{table*}[!t]
    \centering
    \small
    \setlength\tabcolsep{9pt}
    \caption{Analysis of zero-shot generalization capabilities. All frameworks are distilled from SAM-H on 1\% samples of the SA-1B dataset \cite{kirillov2023segment} and evaluated on all samples of each dataset with the box prompt mode. }
    {\scalebox{1}{
    \begin{tabular}{l|cc|ccccccc}
    \hline
     \multirow{2}{*}{Methods} & \#Params & \#FLOPs & \multicolumn{2}{c}{2D} & \multicolumn{2}{c}{3D} & \multicolumn{2}{c}{Video} \\
    \cline{4-9}
    & (Encoder) & (Encoder) &  Dice & mIoU & Dice & HD &  Dice & mIoU \\
    \hline
    SAM-H \cite{kirillov2023segment} & 637.03M & 2733.64G & 84.19 & 74.59 & 89.65 & 25.86 & 91.19 & 84.68\\
    TinySAM \cite{shu2023tinysam} & 6.07M & 36.74G & 78.45 & 65.17 & 75.38 & 51.70 & 70.28 & 56.80 \\
    MobileSAM \cite{zhang2023faster} & 6.07M & 36.74G & 81.73 & 71.69 & 88.76 & 30.39 & 89.92 & 82.89 \\
    RepViT-SAM \cite{wang2024repvit} & 5.12M & 18.61G & 81.84 & 71.83 & 88.83 & 29.79 & 90.09 & 83.06  \\ 
    EfficientSAM \cite{xiong2024efficientsam} & 6.16M & 25.04G & 81.34 & 70.80 & 79.39 & 42.38 & 83.37 & 72.31 \\
    HER-SAM & \textbf{1.57M} & \textbf{6.43G} & \underline{81.98} & \underline{71.97} & \underline{88.92} & \underline{28.15} & \underline{90.17} & \underline{83.26}\\
    \hline
    SAM2-L \cite{ravi2024sam} & 212.70M & 810.99G & 86.76 & 78.03 & 92.23 & 20.38 & 92.88 & 87.06 \\
    HER-SAM2 & \textbf{1.57M} & \textbf{6.43G} & \textbf{85.13} & \textbf{75.61} & \textbf{89.54} & \textbf{26.89} & \textbf{90.47} & \textbf{83.68}\\ 
    \hline
    \end{tabular}}}
    \label{tab4}
\end{table*}

\begin{table*}[!t]
\centering
\small
\setlength\tabcolsep{7.5pt}
\caption{Ablation Study of cascaded token filtering on three high-resolution medical segmentation datasets.}
{\scalebox{1}{
\begin{tabular}{cc|ccc|cc|ccc|cc}
\hline
\multirow{2}{*}{Pre-filtering} & \multirow{2}{*}{Post-filtering} & \multirow{2}{*}{\#Mem} & \multirow{2}{*}{\#FLOPs} & \multirow{2}{*}{\#Params} & \multicolumn{2}{c|}{ISIC-2018} & \multicolumn{3}{c|}{Synapse} & \multicolumn{2}{c}{CVC-ColonDB}\\
\cline{6-12}
& & & & & Dice & mIoU & Dice & mIoU & HD & Dice & mIoU\\
\hline
& & 0.57GB & 9.96G & 3.75M & 82.91 & 75.48 & 79.83 & 73.12 & 52.74 & 76.85 & 69.12 \\
\checkmark & & 0.51GB & 8.76G & 3.39M & 86.17 & 78.95 & 82.56 & 75.89 & 45.18 & 83.27 & 75.95 \\
& \checkmark & 0.65GB & 10.73G & 3.92M & 84.63 & 77.29 & 81.42 & 74.71 & 48.96 & 81.94 & 73.88 \\
\checkmark & \checkmark & 0.59GB & 9.39G & 3.67M & 88.63 & 81.62 & 84.25 & 77.57 & 36.09 & 85.03 & 78.21 \\
\hline
\end{tabular}}}
\label{tab:6}
\end{table*}

\subsection{Ablation Study}
To investigate the effectiveness of our proposed CE-Encoder, DLA, and ME-Decoder, we conduct comprehensive ablation studies across three different high-resolution medical segmentation datasets: ISIC-2018 (2D), Synapse (3D) and CVC-ColonDB (video), as illustrated in Table \ref{tab:5}. We construct the baseline using a standard U-Net architecture without any proposed components, which demonstrates limited performance with 77.12\% Dice on ISIC-2018 and substantial computational overhead of 7.63GB memory and 497.71G FLOPs. By introducing the CE-Encoder with self-attention, the performance achieves significant improvements with Dice increases of 5.33\%, 4.17\% and 3.45\% on ISIC-2018, Synapse and CVC-ColonDB datasets, respectively. The CE-Encoder effectively reduces computational complexity to 324.93G FLOPs while maintaining 5.85GB memory usage. Moreover, we investigate the effect of combining CE-Encoder with DLA, resulting in further performance gains with substantial efficiency improvements, reducing FLOPs to 97.41G. This validates the synergistic effect of our DLA mechanism in selectively retaining important tokens. Finally, we establish the complete HER-Seg framework by incorporating the ME-Decoder with cross-scale segmentation decoding. The complete HER-Seg framework dramatically reduces memory and computational requirements to merely 0.59GB GPU memory, 9.39G FLOPs and 3.67M number of parameters cost with the best performance across all three datasets, achieving Dice increases of 3.87\%, 2.63\% and 5.39\% on ISIC-2018, Synapse and CVC-ColonDB datasets, respectively. These ablation results prove the synergistic contributions of all modules to achieving holistically efficient high-resolution medical segmentation.

\begin{table}[!t]
    \centering
    \setlength\tabcolsep{3pt}
    \caption{Ablation study of cross-scale decoding.}
    {\scalebox{1}{
    \begin{tabular}{ccc|c|cc|ccc|cc}
    \hline
    \multirow{2}{*}{$P_1$} & \multirow{2}{*}{$P_2$} & \multirow{2}{*}{$P_3$} & \multirow{2}{*}{FPS} & \multicolumn{2}{c|}{ISIC-2018} &  \multicolumn{3}{c|}{Synapse} & \multicolumn{2}{c}{CVC-ColonDB}\\
    \cline{5-11}
    &  &   &  & Dice & mIoU & Dice & mIoU & HD & Dice & mIoU\\
    \hline
     &  &  & 41.46 & 86.96 & 79.84 & 81.45 & 75.12 & 42.67 & 81.26 & 74.89   \\ 
    \checkmark &  &  & 41.29 & 87.54 & 80.51 & 82.19 & 75.86 & 40.15 & 82.47 & 76.13   \\ 
    \checkmark & \checkmark & & 41.13 & 88.21 & 81.29 & 83.08 & 76.71 & 39.92 & 83.85 & 77.42   \\
    \checkmark & \checkmark & \checkmark & 40.87 & 88.63 & 81.62 & 84.25 & 77.57 & 36.09 & 85.03 & 78.21  \\
    \hline
    \end{tabular}}}
    \label{tab:8}
\end{table}

\subsection{Zero-shot Generalization Analysis}
To demonstrate the generalization capabilities of our proposed CE-Encoder, we conduct zero-shot segmentation evaluation across all eight high-resolution medical segmentation datasets, including 2D, 3D, and video domains. We follow existing studies \cite{zhang2023faster, xiong2024efficientsam} that leverage 1\% samples of the SA-1B dataset \cite{kirillov2023segment} to conduct feature distillation between CE-Encoder and SAM-ViTs \cite{kirillov2023segment, ravi2024sam} with MSE loss. The pretrained encoder is then combined with the prompt encoder and mask decoder of SAM, called HER-SAM. To show the upper-bound performance of all methods, we adopt the box prompt generated by the same configuration \cite{ma2024segment}.  As shown in Table \ref{tab4}, the original SAM-H \cite{kirillov2023segment} achieves superior performance across all domains but requires substantial computational resources with 637.03M parameters and 2733.64G FLOPs. Recent efficient variants \cite{shu2023tinysam, zhang2023faster} demonstrate significant parameter reduction but suffer from considerable performance degradation. In contrast, HER-SAM demonstrates remarkable efficiency-performance trade-offs, representing a 405.8$\times$ parameter reduction and 425.1$\times$ FLOPs reduction compared to SAM-H. The consistent performance demonstrates that our CE-Encoder preserves the essential segmentation capabilities while dramatically reducing computational requirements. Furthermore, we extend our framework to the latest SAM2 architecture. As with SAM, we distill the knowledge of the image encoder in SAM2-L to our CE-Encoder using the same 1\% samples of the SA-1B dataset. Our HER-SAM2 illustrates remarkable zero-shot capabilities while requiring 135.5$\times$ fewer parameters and 126.1$\times$ fewer FLOPs. These comprehensive evaluations confirm that HER-SAM maintains strong generalization abilities and demonstrates practical applicability for resource-constrained deployment scenarios.

\subsection{Analysis of Efficiency-Performance Trade-offs}
\begin{figure}[!t]
  \centering
  \includegraphics[width=1\linewidth]{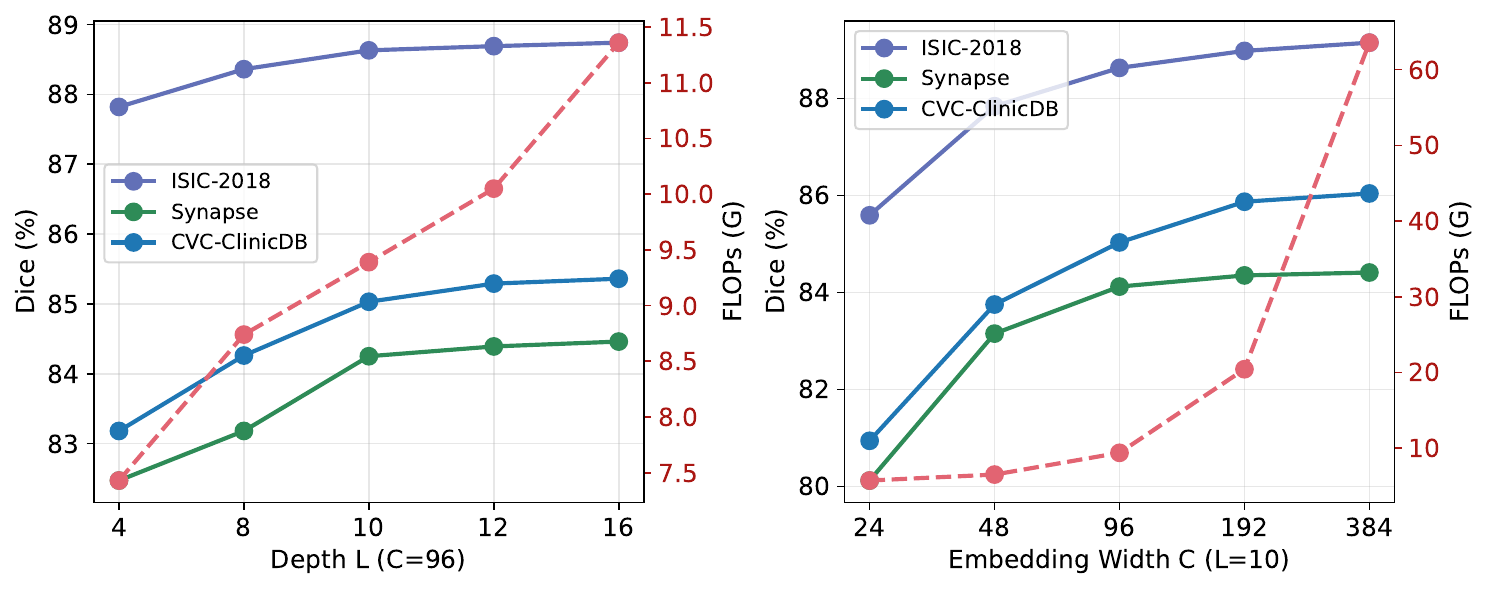}
  \caption{Hyper-parameter analysis of model depth $L$ and embedding width $C$ on the ISIC-2018, Synapse and CVC-ColonDB datasets.}
  \label{fig:ab_token}
\end{figure}

To comprehensively validate the design choices of our CE-Encoder, we conduct extensive ablation studies on the depth $L$ and embedding width $C$ parameters, as detailed in Fig. \ref{fig:ab_token}. We observe that increasing the depth from 4 to 10 layers yields substantial performance improvements across all datasets. Specifically, the Dice metric on Synapse rises dramatically from 80.11 to 84.12. However, further increasing depth to 16 layers provides only marginal gains while significantly increasing computational overhead. Moreover, the extremely narrow embeddings ($C=24$) severely limit model capacity, resulting in poor performance on three high-resolution medical segmentation datasets, increasing $C$ beyond 96 yields diminishing returns. Notably, $C=192$ and $C=384$ achieve only modest improvements at the cost of substantially higher computational complexity. These comprehensive comparison results confirm that the proposed CE-Encoder design with $L=10$ and $C=96$ achieves an optimal balance between segmentation accuracy and computation efficiency.

\subsection{Effectiveness of Cascaded Token Filtering}
To thoroughly evaluate the contribution of our dual-gated linear attention mechanism in DLA, we conduct detailed ablation studies on the token pre-filtering and post-filtering components, as presented in Table \ref{tab:6}. The baseline configuration without any filtering mechanisms achieves moderate performance. When incorporating only the pre-filtering component, we observe substantial improvements across all datasets, with Dice scores increasing by 3.26\%, 2.73\% and 6.42\% on ISIC-2018, Synapse and CVC-ColonDB, respectively, while simultaneously reducing memory consumption to 0.51GB and FLOPs to 8.76G. The post-filtering mechanism alone also contributes positively to segmentation performance with a slight increase in memory and computation cost. Further, DLA combines both pre-filtering and post-filtering to achieve the best performance across all three datasets. These comprehensive evaluations prove the effectiveness of the cascaded token filtering in high-resolution medical segmentation.

\subsection{Significance of Cross-Scale Decoding}
We further conduct ablation studies on the cross-scale decoding, as detailed in Table \ref{tab:8}. The baseline configuration without any multiscale meta-pooling operators achieves 86.96\%, 81.45\% and 81.26\% Dice on ISIC-2018, Synapse and CVC-ColonDB medical segmentation datasets, respectively, while maintaining the highest inference speed of 41.46 FPS. By separately adding Pool $5 \times 5$ ($P_1$), Pool $9 \times 9$ ($P_2$) and Pool $13 \times 13$ ($P_3$) operators, the performance of our HER-Seg is consistently improved with a slight FPS decrease, maintaining the efficiency of inference speed. These comparison results validate the design rationale of our ME-Decoder, demonstrating that cross-scale decoding effectively captures multi-level semantic information without the computational overhead associated with traditional hierarchical architectures, making it highly suitable for high-resolution medical image segmentation tasks and real-time clinical applications.

\section{Conclusion}
In this work, we identify the computational and memory constraints in high-resolution medical image segmentation and propose a holistically efficient framework called HER-Seg to address these critical limitations. The framework integrates a CE-Encoder that utilizes DLA to perform selective token filtering, and a ME-Decoder that leverages cross-scale segmentation decoding to eliminate the demand for hierarchical structures. Extensive experiments on diverse high-resolution medical datasets demonstrate that HER-Seg surpasses state-of-the-art methods with efficient memory and computation cost.

\balance
\bibliographystyle{IEEEtran}
\bibliography{ref}

\end{document}